\documentstyle[11pt]{article}

\newtheorem{theorem}{Theorem}
\newtheorem{conjecture}[theorem]{Conjecture}
\newtheorem{corollary}[theorem]{Corollary}

\newtheorem{lemma}[theorem]{Lemma}

\newtheorem{definition}{Definition}

\def\qed{\hfill $\Box$}

\def\bth{\begin{theorem}}
\def\eth{\end{theorem}}
\def\bc{\begin{corollary}}
\def\ec{\end{corollary}}
\def\bcj{\begin{conjecture}}
\def\ecj{\end{conjecture}}

\title{A randomized algorithm for the on-line weighted bipartite matching problem
\thanks{The authors were partially supported by OTKA grants T034475 and T049398.}}
\author{B\'ela Csaba\thanks{Analysis and Stochastics Research Group of the 
Hungarian Academy of Sciences, Szeged, Aradi v\'ertan\'uk tere 1, H-6720. 
email: {\tt bcsaba@math.u-szeged.hu}} 
\and Andr\'as Pluh\'ar\thanks{Department of Computer Science, 
University of Szeged, Szeged, \'Arp\'ad t\'er 2., H-6720. 
email: {\tt pluhar@inf.u-szeged.hu}}}

\date{}

\begin{document}

\maketitle


\begin{abstract}

We study the on-line minimum weighted bipartite matching problem in 
arbitrary metric spaces. Here, $n$ not necessary disjoint points of a metric space 
$M$ are given, and are to be matched on-line with $n$ points of $M$ revealed 
one by one. The cost of a matching is the sum of the distances of the 
matched points, and the goal is to find or approximate its minimum. 
The competitive ratio of the deterministic problem is known to be 
$\Theta(n)$, see \cite{Kalyanasundaram1, Khuller}.
It was conjectured in 
\cite{Kalyanasundaram2} that a randomized algorithm may perform better 
against an oblivious adversary, namely with an expected competitive ratio
$\Theta(\log n)$. We prove a slightly weaker result by showing a 
$o(\log^3 n)$
upper bound on the expected competitive ratio. As an application the same
upper bound holds for the notoriously hard fire station problem, where $M$ 
is the real line, see \cite{Fuchs, Koutsoupias}.
\end{abstract}




\section{Introduction}

Finding a minimum weight matching in a weighted graph $G$ is a well studied
problem in graph theory. Much less is known about its on-line version; here
we briefly introduce the set-up and the most important results. For more 
thorough references see \cite{Kalyanasundaram1, Kalyanasundaram2, Khuller,
Koutsoupias}.

Let $G$ be an arbitrary weighted graph, and given two players, $A$ and $B$, 
we consider the following {\it on-line matching game} on $G$: First, 
$A$ picks the multiset $S=\{s_1, \dots, s_n\}$ of $V(G)$, these are 
the {\em servers}. Then, one by one, $A$ discloses the {\em requests}, 
again a multiset $R=\{r_1, \dots, r_n\}$ of $V(G)$.
When an element of $R$ is requested, $B$ has to match it with some 
unmatched element from $S$, and $B$ wishes to minimize the cost of the 
resulted matching.

It is clear that usually $B$ cannot reach the offline minimum, and the 
competitive ratio, that is the online cost/offline optimum, is infinite 
if one has no further assumption on $G$ (see Kalyanasundaram and Pruhs, 
and Khuller et al in \cite{Kalyanasundaram1, Khuller}). It was assumed 
in both papers that the weights are nonnegative, and satisfy the triangle 
inequality, so one may refer to the graph $G$ as a metric space 
${M}=(X, d)$ 
with underlying set $X$ and distance function $d$, while the multisets 
$S$ and $R$ are repeated points of $M$.
Then the best competitive ratio is exactly $2k-1$. This is achieved 
for $K_{1, k}$, the so-called star metric space, where the weights are 
all ones.  

The randomized setup for the above on-line game is the following: 
first, $A$ has to construct $S$ and $R$ in advance and disclose $S$.
Then $A$ gives the points of $R$, one by one, but this time he has no 
right to make any changes in the requests, no matter how $B$ is playing.  
That is, not only $R$ but the ordering in which the points of it are 
requested are determined in advance. 
In this setup $B$ has the advantage of using randomness when deciding 
which point of $S$ to be matched with the newly requested point. Let
${\rm opt}(\rho)$ be the total weight of the optimum matching for a sequence 
of requests $\rho$.
We say that $B$'s randomized strategy is $c$--competitive if for every 
request sequence $\rho$ 
$$E[B(\rho)] \le c \cdot {\rm opt}(\rho),$$
\noindent where $E[B(\rho)]$ denotes the expected total weight of the 
matching $B$ finds for $\rho$.
Finding good randomized algorithms for the on-line minimum matching 
problem was first addressed by Kalyanasundaram and Pruhs in 
\cite{Kalyanasundaram2}. They stated that the optimal competitive ratio
for a star metric space is $2H_k-1$, and conjectured an $O(\log n)$
upper bound on the best competitive ratio for arbitrary metric spaces. 
Here and later $n$ stands for the number of servers (or requests). 

Our goal is to show the following theorem.

\begin{theorem} \label{fotetel} There is a randomized on-line weighted 
matching algorithm for arbitrary metric spaces which is 
$O(\log^3 n / \log \log n)$--competitive against an oblivious adversary.
\end{theorem}

The strategy of the proof is the following. First we show that it is 
enough to consider the case when the metric space ${M}$ is a finite
space, indeed $X$ is the set of servers. This will cost only a constant
factor of at most 3.
Then we develop a randomized weighted greedy matching algorithm 
({\bf RWGM}) that has competitive ratio $O(\log n)$ if the points of 
$M$ are the leaves of a 
{\em hierarchically well separated tree}, or {\em HST}. Here the 
distance $d(x, y)$ is defined by adding up the weights on the edges 
of the unique paths connecting $x$ and $y$, and the edge weights 
grow exponentially by the levels of the tree. In our case 
the smallest weights are of size $O(\log n)$.   
In order to use this special case, we recall earlier results on 
probabilistically approximating arbitrary metric spaces by such 
trees next. This approximation contributes
a $O(\log^2 n /\log \log n)$ factor to the competitive ratio, so 
finally we arrive at an algorithm with competitive ratio 
$O(\log^3 n / \log \log n)$. 

Independently of this work Meyerson, Nanavati and 
Poplawski~\cite{MNP} exhibited a randomized on-line algorithm for the matching problem.
They also proved a polylog competitive ratio, and used {\em HST}s.

\section{Discretizing the game} \label{discrete}

Assume that we have an on-line matching algorithm $MA$ that is  
$c$-competitive in the possibly infinite metric space ${M}$ in case 
$R \subset S$ (multiplicities allowed). In this subsection we will 
show that with a small loss in the competitive factor, $MA$ can 
easily be extended to an on-line matching algorithm $MAI$ which 
works for arbitrary $S, R \subset M$. The extension of the algorithm 
is based on a transformation of $R$  which we call {\it discretization}.

Given  $S$ assume that the elements of $R$ appear one after the other. 
For $r_i \in R$ we assign a new point $g(r_i) \in S$.  
We determine $g(r_i)$ in a greedy fashion: 
if $d(s_0,r_i)=\min_{s \in S}{d(s,r_i)}$, then $g(r_i)=s_0$ 
(breaking ties arbitrarily). Clearly, we can find $g(r_i)$ on-line. 
For $s\in S$ denote $rm_s$ the number of requests which are assigned 
to $s$ by $g$. 
The new multiset of requests is called $R'$, in which every $s \in S$ 
appears $rm_s$ times.  $R'$ is the discretized version of $R$.

As above, assume that $MA$ is a $c$-competitive on-line algorithm in the 
case where $R \subset S$. Clearly, after the discretization we arrive at 
an $R'$ such that $R' \subset S$.  
We give another on-line algorithm $MAI$ in the following way: 
we play another, {\it auxiliary} on-line matching game on $M$ using $MA$, 
and use $MA$'s decisions to determine which server $MAI$ chooses to serve 
a request.
Suppose that a request $r \in R$ appears. We determine $g(r)$, and serve 
this request using the server returned by $MA$. If $MA$ chooses $s \in S$ 
to serve $g(r)$, then $MAI$ will serve $r$ using $s$. 

\begin{lemma} \label{diszkret}
If $MA$ is $c$-competitive, then $MAI$ is $(2c+1)$-competitive for 
arbitrary $S, R \subset M$.  
\end{lemma}
 
\noindent {\bf Proof:} We start with some more notation. 
For a matching algorithm $A$ denote $A(r_i)$ the distance from $r_i$ to 
$s$ if $A$ serves this request using $s$. Denote $OM$ the optimal cost 
matching between $S$ and $R$, and let $opt=cost(OM)$. $OM$ induces a 
matching $OM'$ (not necessarily of minimum cost) between $S$ and $R'$ 
in the obvious way: if $(r_i, s_j) \in OM$, then $(g(r_i), s_j) \in OM'$.
For an arbitrary matching $M$, $M(r_i)=d(r_i, s_j)$ if $(r_i, s_j) \in M$.
Finally, let us denote by $opt'$ the total cost of the minimum matching 
between $S$ and $R'$.

From a trivial lower bound on the optimum and by the repeated use of the
triangle inequality we have $\sum_{i=1}^n{d(r_i,g(r_i))} \le opt.$ Note
that $cost(OM') \ge opt'$ by definition. By the 
triangle inequality $MAI(r_i) \le MA(g(r_i)) + d(g(r_i), r_i)$, hence, 
$\sum_{i=1}^{n} MAI(r_i) \le \sum_{i=1}^{n} MA(g(r_i))+opt$.

Again by the triangle inequality, 
$cost(OM'(g(r_i))) \le cost(OM(r_i))+ d(g(r_i),r_i)$ for all $i=1, \dots n$, 
that sum up to $cost(OM') \le cost(OM)+ \sum_{i=1}^n d(g(r_i),r_i)$.
That is
$$opt' \le cost(OM') \le cost(OM)+ \sum_{i=1}^n{d(r_i,g(r_i))} 
\le cost(OM)+opt \le 2 opt.$$
$MA$ is a $c$-competitive on-line algorithm by assumption,
i.\ e., $\sum_{i=1}^n{MA(r_i)} \le c \cdot opt'$.
We know that $MAI(r_i) \le MA(r_i)+opt(r_i)$, therefore,  
$\sum_{i=1}^n{MAI(r_i)} \le c \cdot opt' + opt \le (2c+1) opt.$  \qed 

\smallskip

\noindent {\bf Remark.} Lemma~\ref{diszkret} gives an alternative 
proof of the theorem of Kalyanasundaram and Pruhs \cite{Kalyanasundaram1}, 
that the competitive ratio of the greedy algorithm is at most $2^n-1$. 
Indeed, let $MA$ and $MAI$ be the greedy algorithms for an $n-1$ and an 
$n$ element input, respectively, and use induction.

\section{The algorithm {\bf RWGM}} \label{algoritmus}
Our algorithm, the randomized weighted greedy matching algorithm, 
or {\bf RWGM} algorithm is first developed for special metric spaces.
Assume that the metric space ${M}=(X, d)$ is defined by a weighted tree $T$. 
The set of the leaves of $T$ is $L \subset X$, and the distance $d(x, y)$ 
for the leaves $x, y$ is the sum of the weights on the (unique) path
connecting $x$ and $y$.
Let $\lambda > 1$ be a real number.
\begin{definition} \label{HST}
A $\lambda$--hierarchically well separated tree ($\lambda$--HST) is a rooted 
weighted tree with the following properties:
\begin{itemize}
\item the edge weight from any node to each of its children is the same,
\item the edge weights along any path from the root to a leaf are 
decreasing by the factor $\lambda$ from one level to the next. The weight 
of an edge incident to a leaf is one.
\end{itemize}
\end{definition}
We define the {\bf RWGM} algorithm first, then show in steps that
it is $O(\log{n})$--competitive on a metric space determined by a 
$\lambda$--HST where $\lambda = 2(1+\log{n})$. 

\subsection{{\bf RWGM:} a randomized weighted matching algorithm for 
hierarchically well separated trees}
Let us consider a $\lambda$-HST, denote it by $T=T(V,E,r)$, 
where $V$ is the vertex set, $E$ is the edge set of $T$, and $r$ is
the root. When playing the matching game only leaves of $T$ will be matched 
to leaves of $T$. We denote the set of leaves by $L$. We will
need the notion of a subtree: given $v \in V$, the vertex $u \in V$ belongs 
to the subtree $T_v$ if the only path from $r$ to $u$
contains $v$. Clearly, $T=T_r$, and if $w \in L$, then $T_w$ contains only
the leaf $w$. We have the relation ``$\le$" among the subtrees containing
a certain leaf $w$: $T_u \le T_{u'}$ if $|T_u| \le |T_{u'}|$, 
$w \in T_u$, $w \in T_{u'}$.

\smallskip

In order to get an easier formulation of {\bf RWGM}, we assume that if $u$ 
is a non-leaf vertex of a $(\log n)$-HST, then all of its children
are non-leaves or all are leaves. This can be achieved by inserting ``dummy" 
vertices in the tree. We can also assume that the edge weights on a level 
are equal. (See \cite{Fakcharoenphol}.) 

\smallskip

During the course of satisfaction of the requests, certain vertices will be 
painted green, and leaves may have multiplicities. A vertex $x$ is green if  
the subtree $T_x$ contains at least one unassigned server.
We need multiplicities 
since a point (as a server) may be listed with multiplicity, and also it 
may be requested several times. (Recall from the introduction that $S$ and
$R$ are multisets of $V(G)$.)
The colors and multiplicities of the vertices may change in time, as we 
satisfy the requests and using up the servers. We try to follow
the greedy algorithm, and break ties by random selection by {\em levels}.

Informally, having a request $r$, we try to assign to $r$ a server $s$ as
close as possible according to the tree-metric. One visualizes this as going
up in $T$ until reaching the first green vertex $x$, and then going down to 
an unassigned server. However, going down from $x$ is unintuitive: we choose
uniformly among those edges $(x, y_1), \dots , (x, y_k)$ that lead to 
unassigned servers. One is tempted to go down on $(x, y)$ with probability proportional
to the number of unassigned servers in $T_{y}$. This other approach is analyzed
in~\cite{MNP}.

\medskip

\noindent {\it Formal description of {\bf RWGM}}

\smallskip

In the beginning the adversary $A$ picks leaves of $T$ with
multiplicity, corresponding to the servers $S=s_1, \ldots, s_n$.
(That is if a leaf $x$ is provided $m$ times as a server then 
$x$ has multiplicity $m$.)

We color a vertex $u$ of $T$ green if $T_u$ contains a leaf with 
positive multiplicity, and will call such subtrees green subtrees.

Then $A$ will give us the requests of $R$ one-by-one, denote them 
by $r_1, \ldots, r_n$. 

Set $i=1$.

\begin{itemize}

\item Step 1. The new request is $r_i$. $B$ looks for the smallest 
subtree $T_u$ which contains $r_i$, and $u$ is green.

\smallskip

\item Step 2. Pick a leaf of $T_u$ among the leaves of positive
multiplicity by the algorithm {\bf Pick-a-leaf} with input $u$. 
Let this leaf be $x$, and let $s_i$ (perhaps after reordering) be 
an unused server corresponds to that is matched 
to $r_i$. Decrease the multiplicity of $x$ by one.
\smallskip
\item Step 3. For every green $w \in V$ check whether $T_w$ contains 
a leaf with positive multiplicity. If not, erase $w$'s color.

\smallskip

\item Step 4. If $i \le n-1$, then set $i=i+1$, then go to Step 1.

\item Step 5. If $i=n$, then STOP.

\end{itemize}

\medskip

\noindent Algorithm {\bf Pick-a-leaf}$(u)$

\smallskip

\begin{itemize}

\item Step 1. If the children of $u$ are leaves, then pick randomly, 
uniformly a leaf among those of positive multiplicity. This is 
the leaf we have chosen. STOP.

\item Step 2. If the children of $u$ are not leaves, then denote 
$u_1, u_2, \ldots, u_t$ the green children of $u$.
Pick one randomly, uniformly among them, say, it is $u_i$. Apply 
{\bf Pick-a-leaf}$(u_i)$.

\end{itemize}

\medskip

\begin{theorem}\label{fa}
The algorithm {\bf RWGM} is $O(\log{n})$--competitive on a metric space 
determined by a $\lambda$--HST against an oblivious adversary.
\end{theorem}

\subsection{Proof of Theorem~\ref{fa}}

We prove Theorem~\ref{fa} in steps. First we consider the case 
of uniform metric space where the multiplicities are all ones, but 
the sizes of $S$ and $R$ may not be equal.
Then we discuss the case where $S$ and $R$ have arbitrary multiplicities. 
Finally we prove the general statement for HST's; here the previous
cases provide a basis for induction arguments.

\subsubsection{Uniform case}

In a uniform metric space the distance of two different points is one.
It closely resembles to the star metric space $K_{1, k}$, where the 
leaves are of a distance two from each other. (This explains the extra 
two factor in some of our later formulas.) 

Assume that $U$ is the uniform metric space on $u$ points. 
Let $S= \{s_1, \ldots, s_q\}$ and $R=\{r_1, \ldots, r_t\}$, 
$s_i \not= s_j$ and $r_i\not=r_j$ if $i\not=j$.
We also assume that the points of $R$ are requested in 
increasing order, first $r_1$, then $r_2$, etc., and 
finally $r_t$.

Before dealing with the general case, let us consider a simple but 
instructive example, when $|S|=|R|=q$, and these sets share $q-1$
points. Clearly, the worst case if the first request $r_1$ is not 
in $S$. Assigning $r_1$ to some $s_i$ for which there is an $r_j=s_i$
destroys optimality. This mistake may spread when we match $r_j$.
It was noted in \cite{Kalyanasundaram2} that any randomized on-line
algorithm for that instance has about $\log q$ expected cost, although 
the optimal cost is one. This explains why we have to take special care 
of the common points of $S$ and $R$, and also the order of requests.

\begin{definition}
We say that $s_i \in S$ has a partner if $s_i=r_{j}$ for some 
$r_{j} \in R$.
Similarly, $r_{j} \in R$ has a partner if $s_i=r_{j}$ for some 
$s_{i} \in S$.
\end{definition}

\medskip

We will give an ordering of the points of $S$ using the above mentioned ordering
on $R$. Firstly if there exist $r_j$ and $r_{\ell}$ such that $s_i$ is the 
partner of $r_j$ and 
$s_{k}$ is the partner of $r_{\ell}$ where $j < \ell$, then $s_i <s_{k}$. 
If $s_i$ has a partner 
and $s_{k}$ has no partner, then $s_{i} < s_{k}$ and $r_j < s_k $ for all $j$. 
Finally, we fix an arbitrary ordering among those points of $S$ which 
have no partner in $R$.
Notice, that we can extend the orderings of $S$ and $R$ into an 
ordering ``$<$" of $S \cup R$. This is done such that if $r_i$ is the
partner of $s_j$ then $r_i < s_j$, and for $r_k > r_i$ we have $s_j <r_k$. 
The points of $S$ having no partner
go to the end of the ordering.

Given $r_i \in R$ we associate a weight $w(r_i)$ with it. It is the 
difference 
of the number of servers following, and the number of requests without 
partner preceeding $r_i$. Let us assume that $r_i$ has no partner, then
$$v_i =|\{s_j: s_j > r_i\}| - |\{r_k: r_k < r_i \ 
{\rm and} \ r_k \ {\rm has \ no \ partner}\}|.$$

If $r_i$ has a partner, then let $v_i=0$. Furthermore let
$H_m=1+ \frac{1}{2} +\dots +\frac{1}{m}$, that is the $m^{\rm th}$ 
Harmonic number. Then we define $w(r_i)=H_{v_i}$ (we let $H_f=0$ if
$f \leq 0$).
We need the following useful lemma. 

\begin{lemma} \label{harmonikusok}
For $n \ge 1$, $H_n=1+\frac{H_{n-1}+\ldots+H_{1}}{n}$. 
\end{lemma}

\noindent {\bf Proof:} Trivial computation. \qed

\begin{lemma} \label{uniform}
Let $\delta=|R-S|$. Then in the case above the expected cost of {\bf RWGM} 
is at most $H_q+H_{q-1}+\ldots+H_{q-\delta+1}$. 
\end{lemma}

\noindent {\bf Proof:} We proceed by induction on $q$ that is the size of
$S$. Notice that we may assume that $r_1$ has no partner, otherwise we can 
immediately apply the induction hypothesis. Now $r_1$ is matched to some randomly 
chosen $s_j \in S$.
One can check by the definition of $v_i$ that the weights of the elements of 
$R \setminus \{r_1\}$ are invariant if $s_j$ had no partner. If $s_j$ had 
the partner $r_i$ then the expected new weight of $r_i$ is at most 
$(H_{q-1}+\ldots+H_{1}) /q$. Now by induction one can see that for 
the resulting smaller subproblem the random algorithm has expected cost
$H_{q-1}+\ldots+H_{q-\delta+1}$. Match of $r_1$ to $s_j$ 
costs one, hence, the expected cost of the algorithm is at most
$$1+\frac{H_{q-1}+\ldots+H_{1}}{q} + H_{q-1}+\ldots+H_{q-\delta+1}=
H_q+H_{q-1}+\ldots+H_{q-\delta+1},$$
by Lemma~\ref{harmonikusok}. \qed

\subsubsection{The case of multiplicities}

We want to handle the case when both the servers and the requests
have various multiplicities. Note, that a server with zero multiplicity
simply means that there is no server at that point. If
$U={x_1, \dots, x_u}$, then let $ms(x_i)$ and $mr(x_j)$ are the
multiplicities of servers and requests in point $x_i$ and
$x_j$, respectively. Let $\delta (x_i)=\max\{0, mr(x_i)-ms(x_i)\}$,
$\delta=\sum_{i=1}^u \delta(x_i)$.

\begin{lemma} \label{multiplicitas}
The expected cost of RWGM is at most $H_q+H_{q-1}+\ldots+H_{q-\delta+1}$.
\end{lemma}

\noindent {\bf Proof:} Fix a maximum matching between servers at 
requests which belong to the same point. Pretend that the remaining unmatched
equal servers/requests are at different points, and apply Lemma~\ref{uniform}. \qed

\medskip

\subsubsection{General ${\boldmath \lambda}$-HST trees}
We proceed by induction on the height of the $\lambda$-HST tree. 
First, we need a more technical form of the hypothesis and some
definitions. 

\begin{definition} \label{fordulo}
Given $s \in S$ and $r \in R$, which are matched in
some matching $M$, consider the path connecting them in the HST tree.
Call the point at the highest level of this path the turning point
of $s$ and $r$, shortly $t_M(s, r)$. For a point $u$ of the tree let
$\tau_M(u)$ be the number of $(s, r)$ matched pairs in $M$
for which $u$ is a turning point.
\end{definition}

Given a point $u$, $h(u)$ will denote the height of $u$. We can express 
the cost of an arbitrary matching $M$ as
$$ 2 \sum_{u}{\tau_M(u) \sum_{i=1}^{h(u)}{\lambda^{i-1}}}.$$

Observe that  $\tau_M(u)$ is the same for any optimal matching $M$, hence
in this case we suppress the subscript $M$. Note that $\tau(u)$ is obvious 
to compute.  
Moreover, one can express the optimal cost: 
$${\rm opt}=2 \sum_{u}{\tau(u) \sum_{i=1}^{h(u)}{\lambda^{i-1}}}.$$
For trees of height less than $d$ our induction hypothesis is
the following inequality: 

$${\rm {\bf {E}}[RWGM]} \le 2 \sum_u{\tau(u)\sum_{i=1}^{h(u)} c_i \lambda^i},$$
where $\lambda=2(1+\log n)$, $c_1=1/2$ and $c_t:=c_{t-1}+(1/2)^t$ for $t>1$. 
Notice, that since $c_t \le 1$, the above statement
proves that {\bf RWGM} is $O(\log{n})$-competitive against an oblivious 
adversary implying Theorem~\ref{fa}.

\medskip

For trees of height one the statement follows from Lemma~\ref{multiplicitas} 
and its remark. 
Consider a tree $T$ of height $d$. We make a new tree $T'$ and a new 
instance $S'$ and $R'$. $T'$ comes from $T$ by pruning the leaves,
and for a $u \in T$, $h(u)=1$ we associate the server and request 
multiplicities that of the sum of the server and request multiplicities 
of its descendants in $T$. $T^*$ denotes the set of subtrees of $T$
of height one, i.\ e. the leaves and their parents. Note that we have
to divide the edge weights of $T'$ by $\lambda$ in order to get a 
$\lambda$-HST-tree.

One can cut the optimal cost for $S, R$ and $T$ in two parts.
The first part is the optimal cost of $S', R'$ and $T'$, which we call
${\rm opt}'$. The second
part is the cost incurring on $T^*$, this is ${\rm opt}^*$. 
Here we have to take care of cases when the number of requests
are greater than the number of servers in a subtree $T_u$ $(h(u)=1)$. 
Then we consider the partial optimal matching using those servers.  
Let us call the cost of
this partial matching, ${\rm opt}^*_u$ the optimal for this case. 

Clearly, 
${\rm opt}^*=\sum_{u: h(u)=1}{{\rm opt}^*_u}=\sum_{u: h(u)=1}{2 \tau(u)}$ 
and one concludes that 
$${\rm opt}=\lambda \cdot {\rm opt}'+\sum_{u:h(u)\geq 2}2\tau(u)+{\rm opt}^*.$$ 

Unfortunately, the on-line cost on $T$ is not the sum of the on-line
costs of the two parts if we handle the parts separately, but they are 
closely related. 

For this reason we have to take care of the costs occurring in  
$T^*$ when such a request is assigned to a leaf of a tree $T_u$ which
is not supposed in the optimal matching. The exact form of this statement 
is spelled out in Lemma~\ref{moves}.

Let $\cal M$ be a random matching resulted from the run of {\bf RWGM} on 
our tree. Then $\tau_{\cal M}(u)$ is a random variable for each $u$ 
non-leaf, and $M=\sum_u \tau_{\cal M}(u)$ is a random variable again.

\begin{lemma} \label{moves}
$${\rm{\bf E}}[M] \le \sum_{u: h(u) \geq 1}{\tau(u) \sum_{i=1}^{h(u)}(1+\log{n})^i}.$$
\end{lemma}

\noindent {\bf Proof:} We prove Lemma~\ref{moves} by induction on the 
height of the tree. It is true for trees of height one by 
Lemma~\ref{multiplicitas}. 
Assuming that the lemma is true for trees of height at most $h$, we will 
show it for trees of height $h+1.$ 

Let $T$ be a $\lambda$-HST tree of height $h+1$. We define $T'$ and $T^*$ 
as before. $M'$ is just the truncated sum of $M$ on $T'$. By the induction 
hypothesis we have the following inequality: 

$${\rm {\bf E}}[M'] \le \sum_{u: h(u) \geq 2}{\tau(u)\sum_{i=1}^{h(u)-1}
(1+\log{n})^i}.$$ 
Note furthermore that every extra request arriving from $T'$ to a vertex $u$ 
of height one (i.\ e. to a root of a tree $T_u$ of the forest $T^*$) 
increases the expected cost of {\bf RWGM} in $T_u$ by at most $\log n$ by 
Lemma~\ref{multiplicitas}.

The average cost on the trees of $T^*$ comes from two sources; 
one is $opt^*$, the other is $M'$. In order to get an upper bound on 
the cost on $T^*$ we have to add them up and multiply both by $\log n$, 
according to the explanation in the previous paragraph. This way we have
$${\rm{\bf {E}}}[M] \leq \log n \left\{\sum_{u: h(u)=1}\tau(u)+{\rm{\bf 
{E}}}[M']\right\}+{\rm{\bf {E}}}[M'] =$$
 
$${\sum_{u: h(u)=1}\tau(u)\log{n}+\sum_{u: h(u) \geq 2}{\tau(u)\sum_{i=2}^{h(u)}(1+\log{n})^i} 
\leq \sum_{u: h(u) \geq 1}{\tau(u)\sum_{i=1}^{h(u)}(1+\log{n})^i}},$$
which proves the lemma. \qed

\medskip

Now we will use this lemma to prove that 
$${\rm {\bf {E}}[RWGM] }\le 2 \sum_u{\tau(u)\sum_{i=1}^{h(u)} c_i \lambda^i}.$$

Again we will proceed by induction. Assume that the statement is true for 
trees of height at most $h$, and consider a tree $T$ of height $h+1$. We prune the leaves of
$T$, thereby getting $T'$. Recall, that edge weights in $T$ has to be divided 
by $\lambda$ so as to get that $T'$ is a $\lambda$-HST. 
For $T'$ the statement is true by the induction hypothesis. 
That is, the expected cost of {\bf RWGM} on $T'$ is at most

$${\rm {\bf E}[RWGM}(T')] \le 2 \sum_{u:h(u)\ge 2} \tau(u)
\sum_{i=1}^{h(u)-1}c_i \lambda^i.$$

Clearly, if we add this up with the expected cost at level one, we get an 
upper bound for the expected cost of {\bf RWGM} on $T$:

$${\rm {\bf E}[RWGM}(T)] \le \lambda \cdot {\rm {\bf E}[RWGM}(T')]
+2\cdot {\rm{\bf {E}}}[M].$$

We want to show that this is at most 
$$2\sum_{u}{\tau(u)\sum_{i=1}^{h(u)}c_i\lambda^i}.$$
The coefficient of $\tau(u)$ in the upper bound is less than 
$\sum_{i=1}^{h(u)}c_i\lambda^i$
for every $u$ at level $\ell$. For $\ell=1$, it follows since 
$\log n \leq c_1 2(1+\log n)$.

For $\ell > 1$, we need to show that 

$$\log n \sum_{i=1}^{\ell-1}(1+\log n)^i+ \sum_{i=1}^{\ell-1}c_i
\lambda^{i+1} \leq \sum_{i=1}^{\ell}c_i \lambda^i.$$

It follows if 
$$\log n \sum_{i=1}^{\ell-1}(1+\log n)^i \leq \sum_{i=2}^{\ell}
(c_i-c_{i-1})\lambda^i+c_1\lambda .$$

Since 
$\log n \sum_{i=1}^{\ell-1}(1+\log n)^i=(1+\log n)^{\ell}-(1+\log n)$, 
it reduces to
 
$$(1+\log n)^{\ell} \leq \sum_{i=1}^{\ell}\left(\frac{1}{2}\right)^i
(2+2\log n)^i=\sum_{i=1}^{\ell}(1+\log n)^i.$$ \qed

\section{Approximating by hierarchically well separated trees}

The first results and applications of hierarchically 
well separated trees are due to Bartal, see in \cite{Bartal1, Bartal2}. 
It generalized the earlier works of Karp \cite{Karp} and
Alon et al \cite{Alon} in which they approximated the distances
in certain graphs by using randomly selected spanning trees.

\begin{definition} \label{domination}
A metric space $M=(X,d_M)$ dominates a metric space $N=(X,d_N)$ if 
for every $x,y \in X$ we have $d_N(x,y) \le d_M(x,y)$.
\end{definition}

\begin{definition} \label{approximation}
A set of metric spaces S over $X$ $\alpha$--probabilistically 
approximates a metric space $M$ over $X$, if every metric space
in $S$ dominates $M$, and there exists a probability distribution 
over metric spaces $N \in S$ such that  for every $x,y \in X$
we have $E[d_N(x,y)] \le \alpha d_M(x,y)$.
\end{definition}

The proof of Theorem~\ref{fotetel} is based on the following
theorem.

\begin{theorem}\cite{Fakcharoenphol} \label{loglog}
Every weighted graph on $n$ vertices can be 
$\alpha$--probabilistically approximated by a set of $\lambda$--HSTs,
for an arbitrary $\lambda >1$ where $\alpha=O(\lambda \log{n} /\log \lambda)$.
\end{theorem} 
	
As noted by Bartal \cite{Bartal1}, having an approximation of
a metric space $M$ by HST trees along with a good algorithm for such trees
always results in a good randomized algorithm in that space. So, what we do
is the following. First, preprocessing: given the set of servers $S$, these
points span a sub-metric space ${\cal M_S} \subset {\cal M}$. Clearly, 
$|{\cal M_S}| \le n$, since $S$ is a multiset of $n$ elements. 
We approximate ${\cal M_S}$ 
by a set of $O(\log n)$-HSTs.  Plugging in $\lambda=2(1+\log n)$ into 
Theorem~\ref{loglog} we get there is a probability 
distribution $\cal P$ on these trees such that the expected distortion is 
$O(\log^2 n / \log \log n)$. Choose one tree at random
according to $\cal P$. This finishes the preprocessing. Whenever a request 
$r \in R$ appears, we determine $g(r)$ (see Section~\ref{discrete}), and use 
RWGM with this new request $g(r)$. We proved in Section~\ref{algoritmus}, 
that RWGM is a $O(\log n)$-competitive algorithm in this case. 
Applying Lemma~\ref{diszkret} and Theorem~\ref{loglog}, we get that RWGM is 
$O(\log^3 n / \log \log n)$ competitive for $\cal M$. 
This proves Theorem~\ref{fotetel}. \qed

\medskip

{\bf Acknowledgment.} We thank Endre Szemer\'edi and Kirk Pruhs 
for the fruitful discussions. The numerous advices of anonymous 
referees also improved a lot on the presentation of the paper.


\begin{thebibliography}{999}

\bibitem{Alon} N. Alon, R.\ M. Karp, D. Peleg, D. West,
A graph--theoretic game and its application to the $k$-server problem,
{\em SIAM J. Comput.} {\bf 24} (1) (1995) 78--100.

\bibitem{Bartal1}Y. Bartal,
Probabilistic approximations of metric spaces and its algorithmic 
applications, in: {\em IEEE Symposium on Foundations of Computer Science,} 
1996, pp. 184--193.
\bibitem{Bartal2}
Y. Bartal, On approximating arbitrary metrics by tree metrics, in:
{\em STOC,} 1998.


\bibitem{Bartal3}
Y. Bartal, M. Charikar and R. Raz,
Approximating min-sum k-clustering in metric spaces, 
Thirty-Third Annual ACM Symposium on Theory of Computing, 
pages 11--20, 2001.

\bibitem{Fakcharoenphol}
J. Fakcharoenphol, S. Rao and K. Talwar,
A tight bound on approximating arbitrary metrics by tree metrics,
{\em J. Comput. System Sci.} {\bf 69} (2004), no. 3, 485--497.

\bibitem{Fuchs}
B. Fuchs, W. Hochst\"attler, and W. Kern.
Online matching on a line,
In Hajo Broersma, Ulrich Faigle, Johann Hurink, Stefan Pickl, and Gerhard
Woeginger, editors, {\em Electronic Notes in Discrete Mathematics,} volume
{\bf 13.} Elsevier, 2003.
\bibitem{Kalyanasundaram1}
B. Kalyanasundaram and K. Pruhs,
Online weighted matching,
{\em Journal of Algorithms,} {\bf 14(3)} (1993) 478--488.

\bibitem{Kalyanasundaram2}
B. Kalyanasundaram, K. Pruhs,
On-line network optimization problems, in
{\em Online algorithms: The State of the Art}, eds. A. Fiat and
G. Woeginger, pages 268--280 Lecture Notes in Comput. Sci., 1442,
{\em Springer, Berlin,} (1998)

\bibitem{Kalyanasundaram3}
B. Kalyanasundaram, K. Pruhs,
The online transportation problem,
{\em SIAM J. Discrete Math.} {\bf 13} (2000), no. 3, 370--383.
\bibitem{Karp}
R. Karp,
A 2k-competitive algorithm for the circle.
{\em Manuscript}, August 1989.

\bibitem{Khuller}
S. Khuller, S.\ G. Mitchell, V.\ V. Vazirani,
On-line algorithms for weighted bipartite matching and stable marriages,
{\em Theoret. Comput. Sci.} {\bf 127} (1994), no. 2, 255--267.

\bibitem{Koutsoupias}
E. Koutsoupias, A. Nanavati,
The online matching problem on a line,
{\em Approximation and online algorithms,} 179--191,
Lecture Notes in Comput. Sci., 2909, {\em Springer, Berlin,} 2004.

\bibitem{MNP}
A. Meyerson, A. Nanavati, L. Poplawski,
Randomized on-line algorithms for minimum metric bipartite matching,
SODA (2006).


\bibitem{Tsai}
Y.\ T. Tsai, C.\ Y. Tang, Y.\ Y. Chen,
Average performance of a greedy algorithm for the on-line minimum
matching problem on Euclidean space,
{\em Inform. Process. Lett.} {\bf 51} (1994), no. 6, 275--282.
\end{thebibliography}
\end{document}